\documentstyle[epsfig]{elsart}
\begin{document}
\def\prl{{\em Phys. Rev. Lett. }}
\def\prc{{\em Phys. Rev. {\bf C} }}
\def\jap{{\em J. Appl. Phys. }}
\def\ajp{{\em Am. J. Phys. }}
\def\nima{{\em Nucl. Instr. and Meth. Phys. {\bf A} }}
\def\npa{{\em Nucl. Phys. {\bf A}}}
\def\npb{{\em Nucl. Phys. {\bf B}}}
\def\epjc{{\em Eur. Phys. J. {\bf C}}}
\def\plb{{\em Phys. Lett. {\bf B}}}
\def\mpla{{\em Mod. Phys. Lett. {\bf A}}}
\def\pr{{\em Phys. Rep.}}
\def\zpc{{\em Z. Phys. {\bf C}}}
\def\zpa{{\em Z. Phys. {\bf A}}}
\begin{frontmatter}
\title{Transverse Energy per Charged Particle and Freeze-Out Criteria \\
in Heavy-Ion Collisions}
\author[UCT]{J.~Cleymans},
\author[UCT,IOP]{R.~Sahoo},
\author[IOP]{D.P.~Mahapatra},
\author[VECC]{D.K. Srivastava},
\author[UCT]{S.~Wheaton}
\address[UCT]{UCT-CERN Research Centre and Department  of  Physics,
University of Cape Town, Rondebosch 7701, South Africa}
\address[IOP]{Institute of Physics, Sachivalaya Marg, Bhubaneswar 751005, India}
\address[VECC]{Variable Energy Cyclotron Centre, 1/AF Bidhan Nagar, Kolkata 700064, India}
\date{\today}
\newcommand{\eovern}{\left<E\right>/\left<N\right>}
\begin{abstract}
In relativistic nucleus-nucleus collisions the  transverse energy per charged particle, $E_T/N_{\textrm{ch}}$,
increases rapidly with beam energy and 
remains approximately constant at about 800 MeV for beam energies 
from SPS to RHIC. 
It is shown that the hadron resonance gas model describes the energy 
dependence,
as well as the lack of centrality dependence, qualitatively.
The  values of  $E_T/N_{\textrm{ch}}$ are related to the 
chemical freeze-out criterium $E/N\approx 1$ GeV valid for primordial hadrons. 
\end{abstract}
\end{frontmatter}
\section{Introduction}

 In this paper we investigate the  
transverse energy per charged particle, $E_T/N_{\textrm{ch}}$,
for  beam energies ranging from about 1 AGeV up to 200 AGeV. 
In this energy range,
$E_T/N_{\textrm{ch}}$ at first increases rapidly from SIS~\cite{fopi}
  to AGS~\cite{e802,e814_6GeV}, then saturates to 
a value of about 800 MeV at SPS~\cite{wa98_17GeV,na49_17GeV,na49} energies 
and remains constant
up to the highest available RHIC energies~\cite{star200GeV,phenix130GeV,phenixSyst}. 
The present analysis of $E_T/N_{\textrm{ch}}$ 
uses the hadron resonance gas model (thermal model) which describes 
the final state in relativistic heavy-ion collisions as composed of 
hadrons, including heavy hadronic resonances
as being  in thermal and chemical equilibrium.
It has been known for many years~\cite{UnifiedFO} that the chemical freeze-out can be 
described by the condition $E/N \approx $ 1 GeV, where $E$ and $N$ are, 
respectively the total energy and  particle number of the 
primordial hadronic resonances before they decay into stable hadrons.
This quantity cannot be determined directly from experiment unless the 
final state multiplicity is  low and hadronic resonances can be identified,
which is not the case in  relativistic heavy-ion collisions.
Our analysis therefore starts by relating the number of charged particles 
seen in the detector to the number of primordial hadronic resonances and 
the transverse energy to the energy $E$ of primordial hadrons.
In this paper all thermal model calculations were performed using the 
THERMUS package~\cite{THERMUS}. 
\section{Dependence of $E_T/N_{\textrm{ch}}$ on Beam Energy and Centrality}
The transverse energy, $E_T$,  is defined as the energy deposited transverse to the beam
direction in a given interval of pseudo-rapidity $\eta$.
The transverse energy  has two components, the hadronic one,
$E_T^{\textrm{had}}$,  and the electromagnetic one, $E_T^{\textrm{em}}$,
coming from the electromagnetic particles (photons, electrons and positrons).
Electromagnetic calorimeters are used to measure $E_T^{\textrm{em}}$ 
whereas hadronic
calorimeters or the Time Projection Chamber (for particle identification
and momentum information) are used to measure $E_T^{\textrm{had}}$.
The energy of a particle is defined as being the kinetic energy for nucleons, 
for anti-nucleons 
as the total energy plus the rest mass and for all other particles as the total
energy~\cite{star200GeV,phenix130GeV,helios}. 

In the experiments 
the transverse energy and the 
charged particle multiplicity are measured in a similar way so that most of the systematic 
uncertainties cancel out in the ratio. 
Experiments  
have reported a constant value of the ratio  
$E_T/N_{\textrm{ch}}~\sim~0.8$ GeV from 
SPS to RHIC~\cite{star200GeV,phenixSyst}, 
with the ratio being almost independent of centrality of the 
collision for all measurements at different energies. 
In all cases the value of $E_T/N_{\textrm{ch}}$ has been taken for the 
most central collisions, at the end of this paper we consider 
the centraly dependence of  $E_T/N_{\textrm{ch}}$. 
When this ratio is observed for the full range of center of mass energies, 
it shows two regions~\cite{phenixSyst}. 
In the first region from lowest $\sqrt{s_{NN}}$ 
to SPS energy, there is a steep increase of the $E_T/N_{\textrm{ch}}$ ratio with $\sqrt{s_{NN}}$. 
In this regime, the increase of $\sqrt{s_{NN}}$ causes an increase in the $\left<m_T\right>$
 of the 
produced particles. In the second region,  SPS  to higher 
energies, the $E_T/N_{\textrm{ch}}$ ratio is very weakly dependent on 
$\sqrt{s_{NN}}$. 
The energy
pumped into the system by the increase of $\sqrt{s_{NN}}$ is converted mainly into particle
production.
 This observation is quite remarkable and requires the help of models for a better 
understanding of the underlying 
physics. 

\begin{figure}
\begin{center}
\includegraphics[width=5.5in]{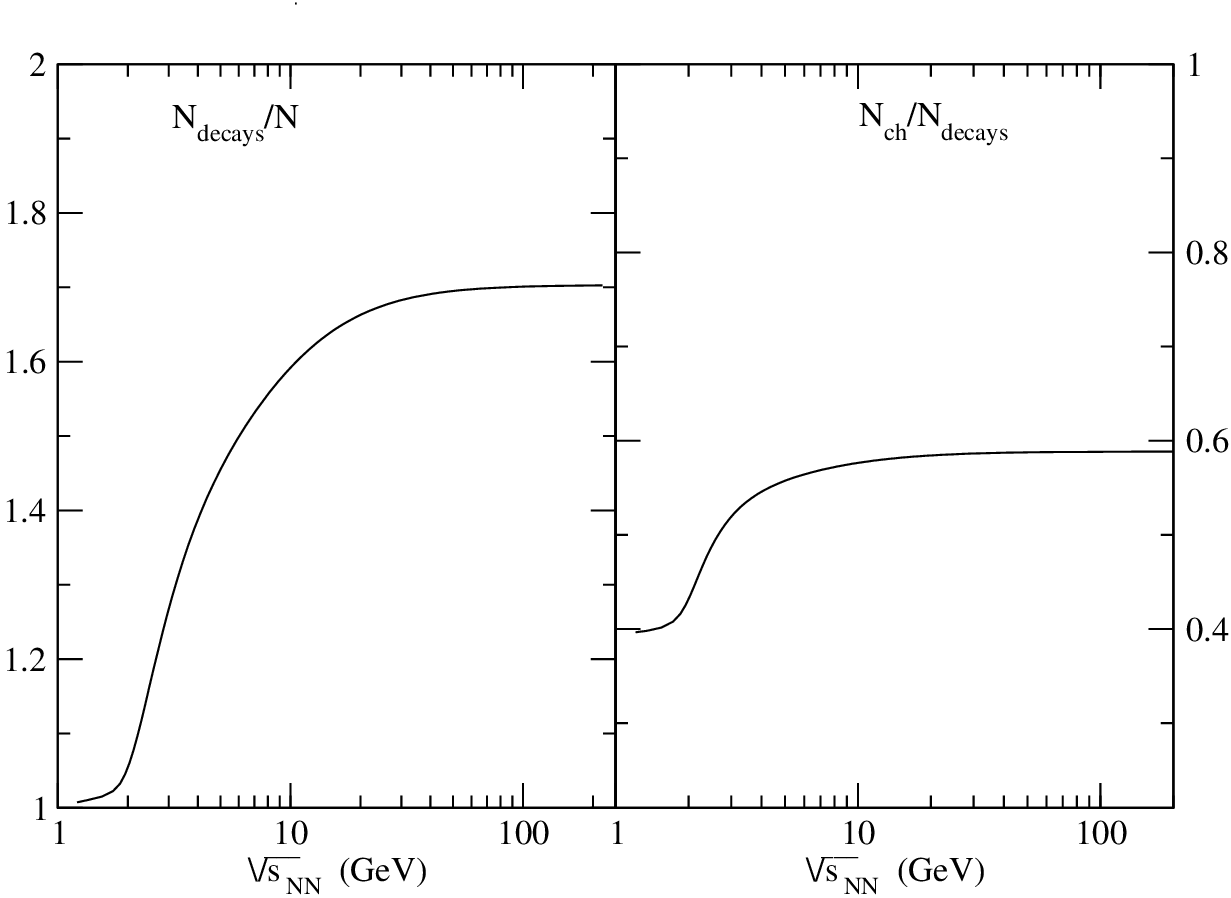}%
\caption{The variation of $N_{\textrm{decays}}/N$ (left) and 
$N_{\textrm{ch}}/N_{\textrm{decays}}$ (right) 
with $\sqrt{s_{NN}}$ in the thermal model with $E/N$ = 1.08 GeV as freeze-out criterium.
}
\label{nCh}
\end{center}
\end{figure}

To estimate  $E_T/N_{\textrm{ch}}$ in the thermal model 
we relate the number of charged particles,  $N_{\textrm{ch}}$,
to the number, $N$, of primordial hadrons.
To estimate the charged particle multiplicity at different center of mass energies from 
the thermal model, we proceed as follows. First we study the variation of the
ratio of the total particle multiplicity in the final state, $N_{\textrm{decays}}$, 
and that in the primordial
i.e. $N_{\textrm{decays}}/N$ 
with $\sqrt{s_{NN}}$.
This ratio starts from one, since there are only very few 
resonances produced at low beam energy 
and becomes almost independent of energy after SPS energy.
The value of $N_{\textrm{decays}}/N$ in the region where it is 
independent of $\sqrt{s_{NN}}$
is around 1.7. The excitation function of 
$N_{\textrm{decays}}/N$
is shown in Fig.~\ref{nCh}(a). Secondly, we have studied the variation of the ratio of
charge particle multiplicity and the particle multiplicity in the final state
($N_{\textrm{ch}}/N_{\textrm{decays}}$) with $\sqrt{s_{NN}}$. 
This is shown in Fig.~\ref{nCh}(b). 
The $N_{\textrm{ch}}/N_{\textrm{decays}}$ ratio starts around 0.4 at
lower $\sqrt{s_{NN}}$ and shows an energy independence at SPS and higher energies.
At lower SIS energy, the baryon dominance at mid-rapidity makes 
$N_{\textrm{ch}}/N_{\textrm{decays}}\sim N_{\textrm{proton}}/N_{\textrm{(proton+neutron)}}$ which 
has a value of 0.45 for  Au-Au collisons

As the next step we connect the transverse energy $E_T$ to the the energy of the primordial
hadrons $E$.  We start by relating the two quantities for a static fireball.
In this case  one has
\begin{equation}
\left<E\sin\theta\right> = V \int {d^3p\over (2\pi)^3 } E\sin\theta~~f(E)
\end{equation}
where $f(E)$ is the statistical distribution factor, e.g. for a Boltzmann distribution
it is given by $f(E) = \exp\left(-E/T\right)$. It is straightforward to re-write this 
expression as 
\begin{eqnarray}
\left<E\sin\theta\right> &=&V {\pi\over 4}\int {d^3p\over (2\pi)^3 } E~f(E)\nonumber\\
                         &=& {\pi\over 4}\left<E\right>
\end{eqnarray}
Thus, for a static fireball, the 
transverse energy is related to the total energy by a simple factor  of $\pi/4$.
In the hadronic resonance gas model there is a sum over all hadrons; furthermore, taking
into account the experimental configuration which leads to adding 
the mass of the nucleon for anti-nucleons and subtracting the same for
nucleons one has
\begin{eqnarray}
\left<E_T\right> & \equiv & V\sum_{i={\textrm{Nucleons}}}\int\frac{d^3p}{(2\pi)^3} (E_i-m_N)\sin\theta~~f(E_i) \nonumber\\
&&+V\sum_{i=\textrm{Anti-nucleons}}\int 
\frac{d^3p}{(2\pi)^3} (E_i+m_N) \sin \theta~~f(E_i)\nonumber\\
&&+V\sum_{i=\textrm{All ~Others}}\int \frac{d^3p}{(2\pi)^3} E_i \sin \theta~~f(E_i) , \nonumber \\
&= &  \frac{\pi}{4}\left[\left<E\right>-m_N\left<N_B-N_{\bar B}\right>\right] .
\end{eqnarray}

The above equation relates the transverse energy measured from the data and that
estimated from the thermal model. 
In the limit of large beam energies one has
\begin{eqnarray}
\lim_{\sqrt{s_{NN}}\rightarrow\infty}{\left<E_T\right>\over N_{\textrm{ch}}} 
&=&{\left<E_T\right>\over 0.6 N_{\textrm{decay}}}\nonumber\\
&=&{\pi\over 4}{1\over 0.6}{E\over 1.7 N}\approx 0.83 ,
\end{eqnarray}
which is close to the value measured at RHIC.
The measured $E_T$
will be affected by the radial flow 
and by the difference between chemical freeze-out and kinetic freeze-out temperatures; 
these effects
will lead to corrections  which tend to largely cancel each other.
A detailed comparison of this, in the framework of a single freeze-out temperature model
and limited to RHIC energies, has been made in Ref.~\cite{prorok}.

\begin{figure}[t]
\begin{center}
\includegraphics[width=4in]{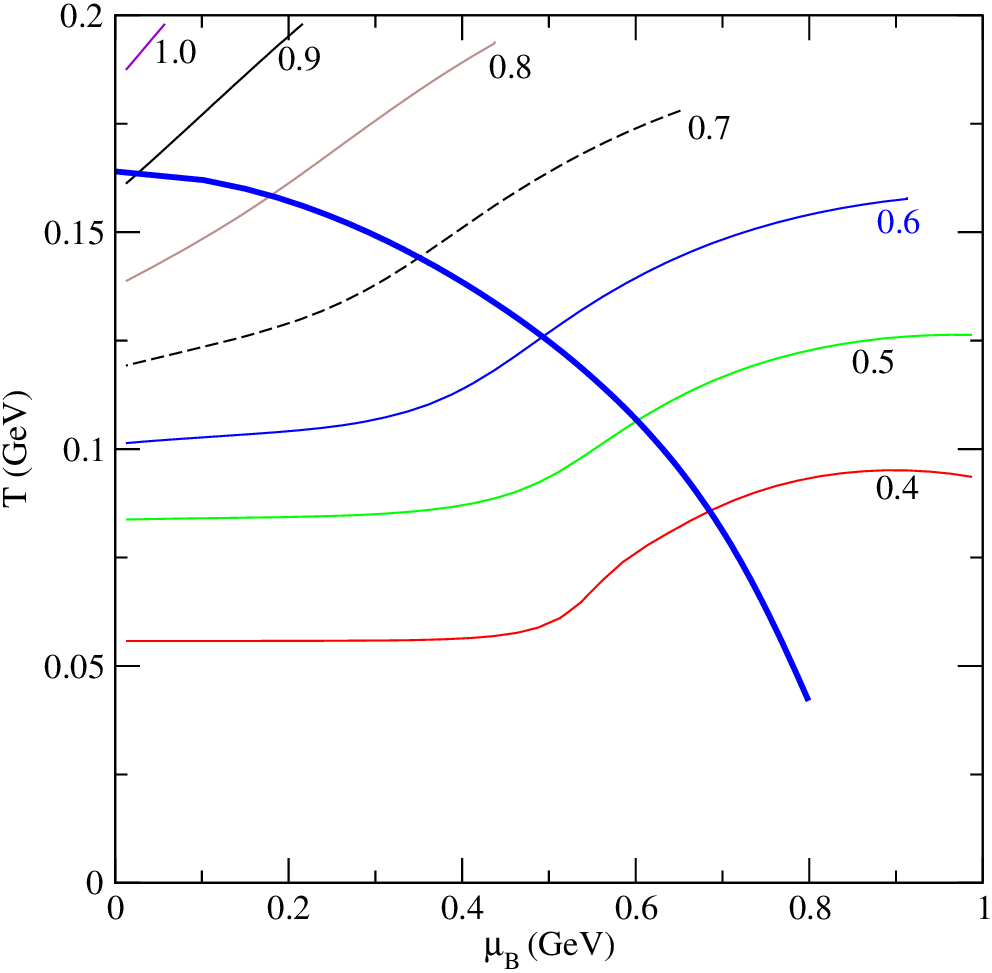}
\caption{Lines of constant $E_T/N_{\textrm{ch}}$ from thermal model without flow
are shown in the $(T,\mu_B)$ plane. The chemical freeze-out condition of 
$E/N = 1.08$ GeV is also shown.}
\label{etNchT}
\end{center}
\end{figure}

\begin{figure}
\begin{center}
\includegraphics[width=5.5in]{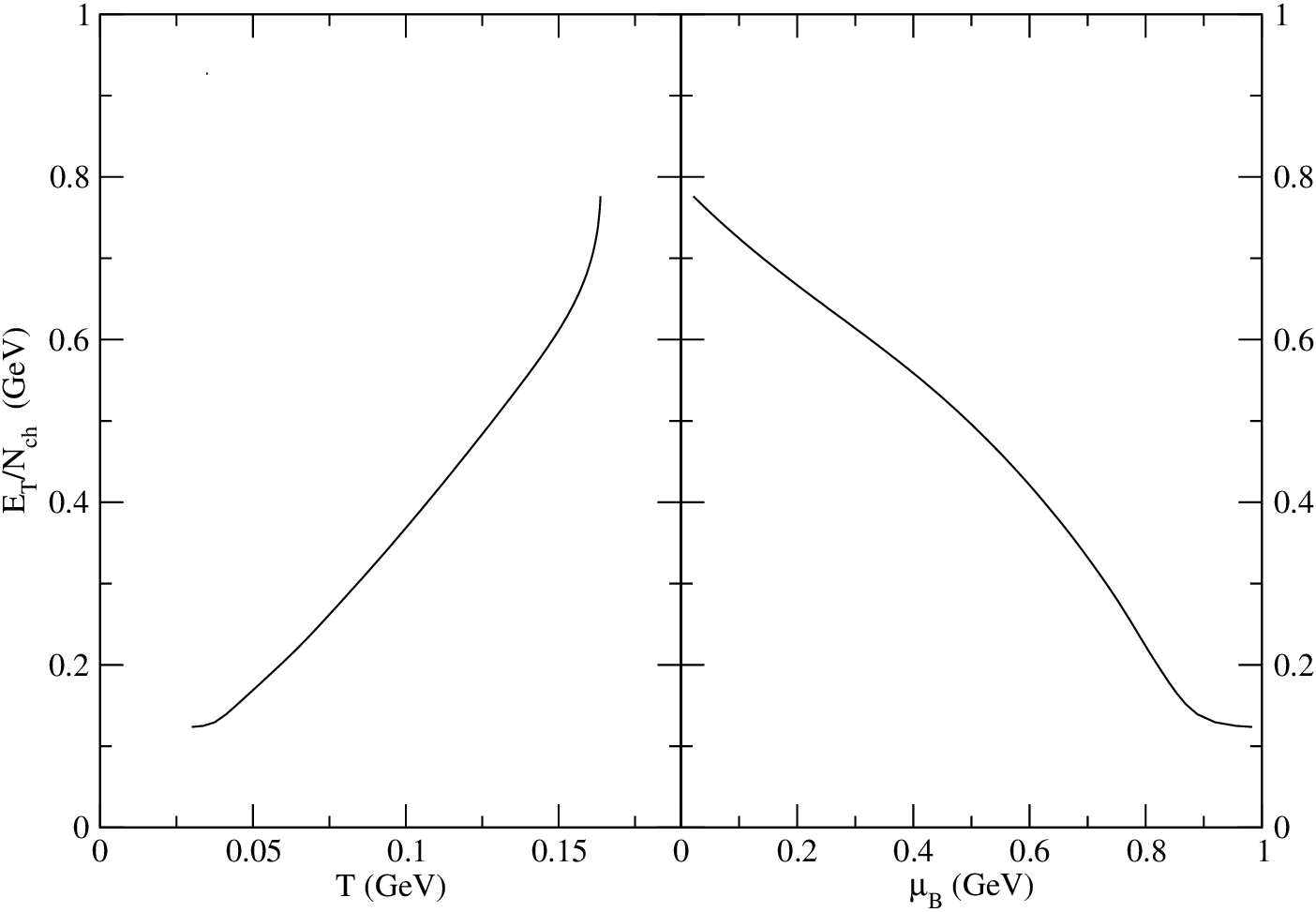}
\caption{The variation of $E_T/N_{\textrm{ch}}$ with $T$ (left) and 
the variation of $E_T/N_{\textrm{ch}}$ with $\mu_B$ (right)}
\label{etNchTMuB}
\end{center}
\end{figure}

In Fig.~\ref{etNchT} we  plot lines of constant $E_T/N_{\textrm{ch}}$
in the  $(T,\mu_B)$-diagram.
For low values of  $E_T/N_{\textrm{ch}}$, these lines are almost indepent of 
$\mu_B$, this is mainly a consequence of subtracting $m_N$ in the definition of
$E_T$, thus taking away much of the influence of nucleons. Only towards larger
values of $\mu_B$  there is a notable dependence on this variable.
To compare with the chemical freeze-out condition, we show also
the chemical freeze-out curve
 in the same  plane (Fig.~\ref{etNchT}). 
At higher 
energies, when $\mu_B$ nearly goes to 
zero, the transverse energy production is mainly due to the meson content in the matter.
The intersection points of lines of constant $E_T/N_{\textrm{ch}}$ and 
the freeze-out line give the values of $E_T/N_{\textrm{ch}}$ at the chemical freeze-out. 
Hence at freeze-out, given the values of $E_T/N_{\textrm{ch}}$ from the experimental 
measurements we can determine $T$ and $\mu_B$ of the system.
In Fig.~\ref{etNchTMuB}, we plot  the ratio $E_T/N_{\textrm{ch}}$ 
as a function of the temperature $T$ and as a function of $\mu_B$.
It can be seen that the relation between the $E_T/N_{\textrm{ch}}$ 
and $T$ is linear
to a good approximation, similarly for the relation with $\mu_B$.

\begin{figure}
\begin{center}
\includegraphics[width=5.5in]{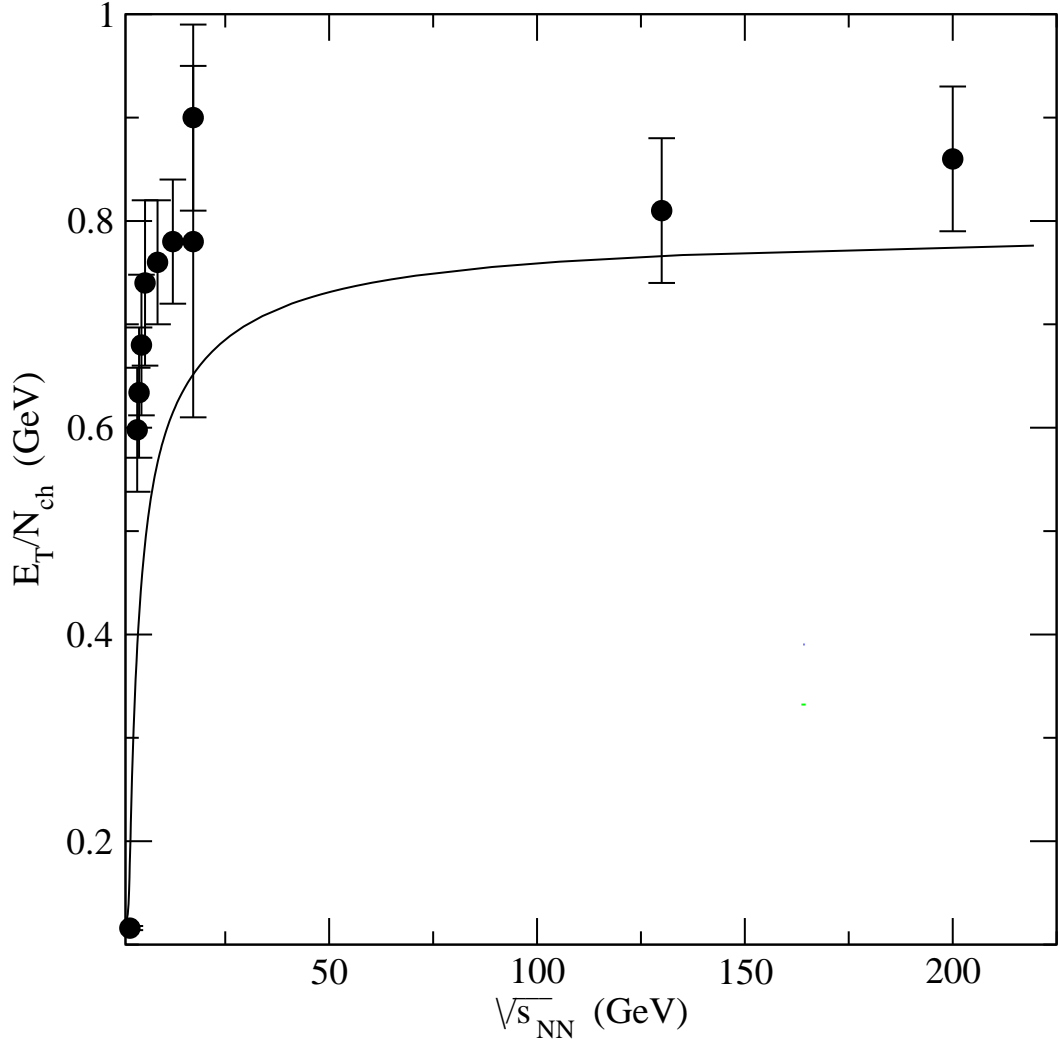}
\caption{Comparison between experimental data for
 $E_T/N_{\textrm{ch}}$ with $\sqrt{s_{NN}}$ and the thermal model
using  $E/N = 1.08$ GeV as  the freeze-out condition.}
\label{etNchCM}
\end{center}
\end{figure}

\begin{figure}
\begin{center}
\includegraphics[width=5.5in]{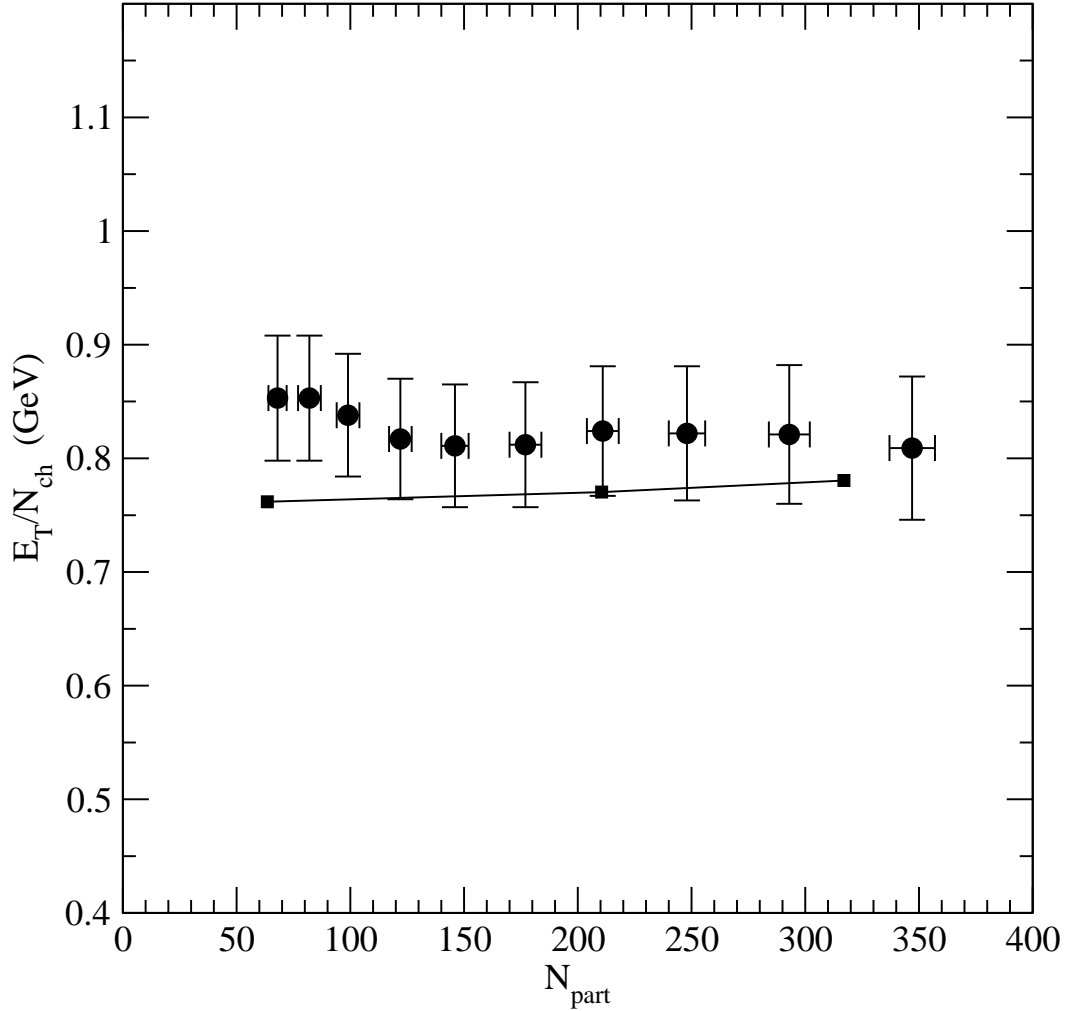}
\caption{The variation of $E_T/N_{\textrm{ch}}$ with $N_{\textrm{part}}$ for 130 GeV Au+Au 
collisions at RHIC with the corresponding thermal model estimate.}
\label{etNchNp}
\end{center}
\end{figure}

For the most central collisions, the variation of $E_T/N_{\textrm{ch}}$ with center of mass energy
is shown in Fig.~\ref{etNchCM}. The data have been taken from 
Ref.~\cite{fopi,e802,e814_6GeV,wa98_17GeV,na49_17GeV,na49,star200GeV,phenix130GeV,phenixSyst},
and are compared with the corresponding calculation
from the thermal model with chemical freeze-out.
We have checked explicitly
that other freeze-out criteria discussed in the literature give almost identical results for the behavior of 
$E_T/N_{\textrm{ch}}$ as a function of $\sqrt{s_{NN}}$;
this is the case for the fixed baryon plus anti-baryon density
condition~\cite{PBMS}
and also for fixed normalised entropy density condition,
$s/T^3$ = 7~\cite{Majiec,TawfikEntropy,tawfik2}.
It has been observed from SPS to RHIC \cite{star200GeV,phenixSyst}, that the ratio   
 $E_T/N_{\textrm{ch}}$, is almost independent of the centrality
of the collisions which is represented by the number of participant nucleons. 
To understand the variation of $E_T/N_{\textrm{ch}}$ with collision centrality, we 
have estimated $E_T/N_{\textrm{ch}}$ for 130 GeV Au+Au collisions at RHIC, for different 
centrality classes~\cite{cfNp}. This is compared with the corresponding data in 
Fig.~\ref{etNchNp}. The effect of flow is not taken into account in the model 
calculations for the centrality behavior. The model agrees  well 
with the experimental data 
for the centrality behavior. Again, we have checked explicitly that other
freeze-out criteria lead to similar results
~\cite{PBMS,Majiec,TawfikEntropy,tawfik2}.
\section{Summary}
In conclusion, we have discussed the connection between  $E_T/N_{\textrm{ch}}$  
and the ratio of primordial energy to primordial particle 
multiplicity, $E/N$,  from the thermal model.  
This
model, when  combined with  chemical freeze-out criteria 
explains the data over all available measurements for the $\sqrt{s_{NN}}$ 
behavior of $E_T/N_{\textrm{ch}}$. 
It has to be noted that  variables like
$E_T/N_{\textrm{ch}}$, the chemical freeze-out temperature $T_{\textrm{ch}}$, 
$N_{\textrm{decays}}/N_{\textrm{primordial}}$ 
and $N_{\textrm{ch}}/N_{\textrm{decays}}$ discussed
in this paper, show  saturation starting at SPS 
and continuing to higher center of mass energies. 
This observation along with the centrality independence of
$E_T/N_{\textrm{ch}}$ is not inconsistent with the simultaneity
of chemical and kinetic freeze-out at higher energies~\cite{sFO}.

\section{Acknowledgement}
Three of us (JC,~RS,~DKS) would like to acknowledge the financial support of
the South Africa-India Science and Technology agreement.

\end{document}